\documentclass{osa-article}

\journal{oe}

\articletype{Research Article}

\usepackage{siunitx}

\begin{document}
\title{High performance superconducting nanowire single photon detectors operating at temperature from 4 to 7 K}
\author{Ronan Gourgues,\authormark{1*} Johannes W. N. Los,\authormark{1} Julien Zichi,\authormark{2} Jin Chang,\authormark{3}  Nima Kalhor,\authormark{1} Gabriele Bulgarini,\authormark{1}Sander N. Dorenbos,\authormark{1} Val Zwiller,\authormark{2} and Iman Esmaeil Zadeh\authormark{3}}

\address{\authormark{1}Single Quantum B.V., 2628 CH Delft, The Netherlands\\
\authormark{2}Department of Applied Physics, Royal Institute of Technology (KTH), SE-106 91 Stockholm, Sweden\\
\authormark{3}Optics Research Group, ImPhys Department, Faculty of Applied Sciences, Delft University of Technology, Lorentzweg 1, 2628 CJ Delft, The Netherlands
}

\email{\authormark{*}ronan@singlequantum.com}


\begin{abstract} 
We experimentally investigate the performance of NbTiN superconducting nanowire single photon detectors above the base temperature of a conventional Gifford-McMahon cryocooler (\SI{2.5}{K}). By tailoring design and thickness (8 - 13 nm) of the detectors, high performance, high operating temperature, single-photon detection from the visible to telecom wavelengths are demonstrated. At \SI{4.3}{K}, a detection efficiency of 82 $\%$ at 785 nm wavelength and a timing jitter of 30 $\pm$ 0.3 ps are achieved. In addition, for 1550 nm  and similar operating temperature we measured a detection efficiency as high as 64 $\%$. Finally, we show that at temperatures up to \SI{7}{K}, unity internal efficiency is maintained for the visible spectrum. Our work is particularly important to allow for the large scale implementation of superconducting single photon detectors in combination with heat sources such as free-space optical windows, cryogenic electronics, microwave sources and active optical components for complex quantum optical experiments and bio-imaging. 
\end{abstract}

\section{Introduction}
Since Superconducting Nanowire Single Photon Detectors (SNSPDs) were introduced in 2001 \cite{doi:10.1063/1.1388868}, their performance has been dramatically improved to become the leading technology in single photon detection. Nowadays, SNSPDs offer outstanding performance, with system detection efficiencies of 80 - 95 $\%$, from the visible to the infrared  \cite{Wollman:17,Wang:19,doi:10.1063/1.5000001,Marsili2013}, dark count rates in the mHz range  \cite{Schuck2013a,Marsili2013}, detection count rates up to 1.5 GHz  \cite{8627992}, and timing jitter as low as < 10 ps \cite{2018arXiv180106574E,Korzh:18,PhysRevB.98.134504}. High performance SNSPDs operating at \SI{4}{K} and above, are particularly interesting due to the vast number of potential applications. It has been proposed that for superconducting quantum computing, part of the qubit readout can be performed with optical signals within the dilution refrigerator. Due to limited cooling power available at the quantum processor stage (below \SI{1}{K}), the readout device has to be located at a higher temperature stage of the cryostat (\SI{4}{K}) \cite{8036394}. Another potential application is optical quantum information processing, which is based on efficient generation and detection of on demand entangled single photons with effective electronic control of the single photon source spin. Promising candidates are solid state emitters such as quantum dots (QDs) and nitrogen-vacancy (NV) centres in diamond which emit in the range of 600 - 1550 nm \cite{Montinaro2014,Chen2018,Manson_2018,doi:10.1021/acs.nanolett.8b00550}. To achieve scalability and more efficient photon extraction \cite{Zadeh2016,PhysRevX.5.031009,Gao2015}, it is required to integrate single photon sources and single photon detectors with photonic circuits \cite{Gourgues:19}. Extending the complexity of integrated photonic circuit with active optical components such as modulators \cite{Gehl:17} will increase the base temperature of the cryostat due to the heat load of the wiring connecting the components with external room temperature control electronics. Thus, this will have a detrimental effect on the performance of the SNSPD. Currently, SNSPDs require sophisticated cooling systems with sizeable power consumption (>\SI{1}{kW}) which restricts their use mainly to the research community. Several high critical temperature (Tc) superconducting materials, such as magnesium diboride (MgB$_{2}$) have been investigated to detect single photons but efficient detectors have yet to be demonstrated \cite{Shibata_2013}. On the other hand, semiconductor avalanche photodiodes (APDs) are an alternative technology in the VIS - NIR range, operating at room temperature and being relatively inexpensive. However, APDs suffer from after-pulsing, higher dark count rates, long non-Gaussian instrument response function and thus limited time resolution \cite{doi:10.1063/1.2221516}. Recently, a low power consumption and compact closed-cycle cooling cryostat suitable for space applications has been developed and engineered by different international research laboratories \cite{Gemmell_2017,8657757,655397499}. For such cryocoolers, the base temperature reached is $\sim$ \SI{4}{K} which drastically limits performance of existing SNSPDs.
\\
In this work, we engineer NbTiN SNSPDs to achieve high efficiency and high time resolution at \SI{4}{K} and above, paving the way for integration of SNSPDs in complex, large scale and portable optical experiments.
\section{SNSPDs fabrication}
\begin{figure}[hbtp!]
      \centering
      \includegraphics[width=11.5 cm]{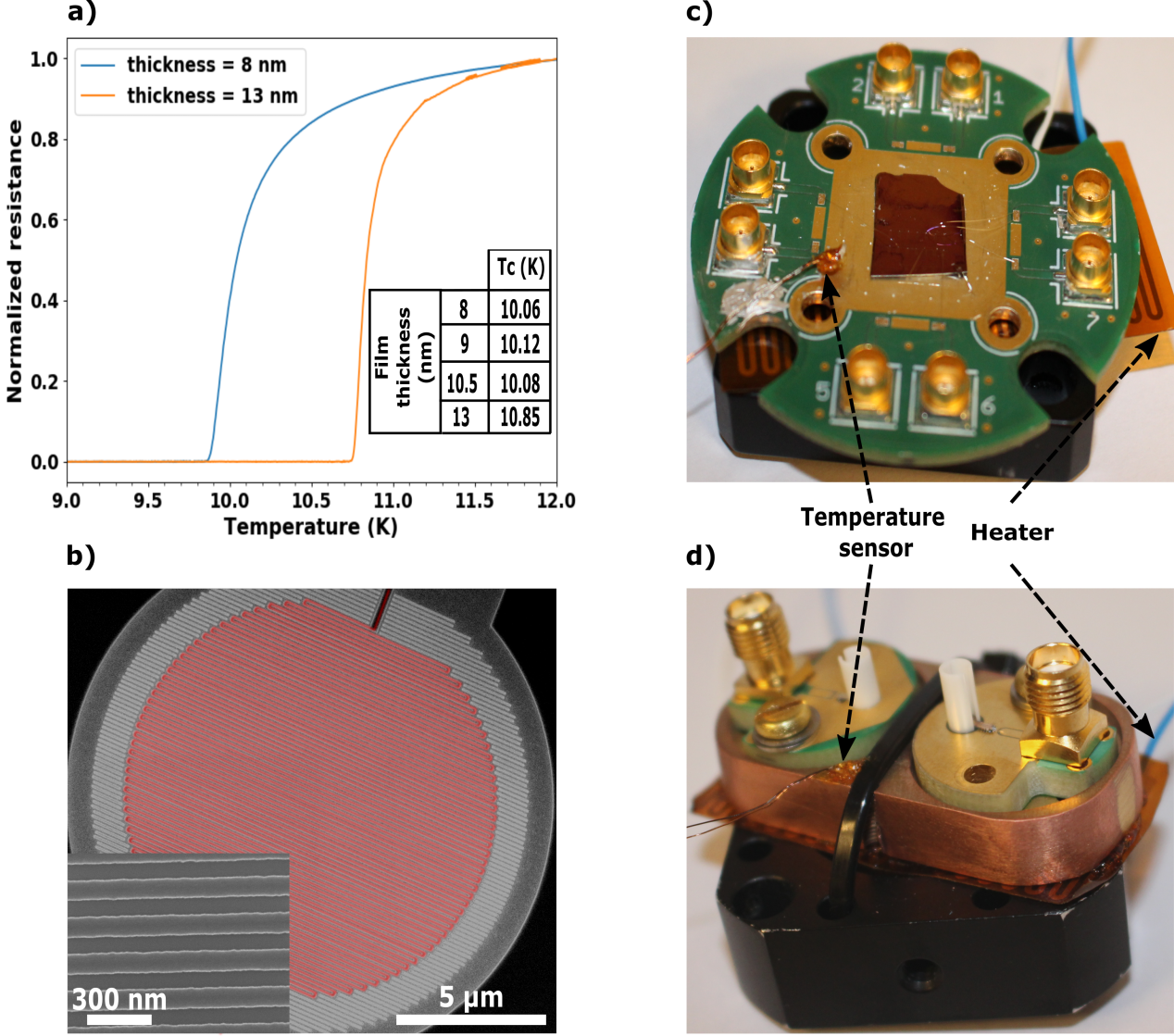}
      \caption{(a) Measurements of the superconducting transition temperature for 8 nm and 13 nm film thickness. The table in the inset summarizes the Tc for the studied films. The error is estimated to be $\pm$ \SI{0.1}{K}. 
		       (b) False-color SEM image of the superconducting single photon detector meander. The inset shows the 70 nm nanowire width of the superconducting device.   
		       (c) Free-space holder with the sample wire-bonded in the center of the PCB.
		       (d) Fiber-coupled superconducting detector wire-bonded to the PCB, mounted on the oxygen free copper block.}
	\label{fig:fig1}
\end{figure}
The first fabrication step of the SNSPD is the deposition of NbTiN films by magnetron co-sputtering in an Ar and N$_{2}$ atmosphere on a Distributed Bragg Reflector (DBR) mirror optimized for the wavelength of interest. The sputtering recipe was, in a separate study \cite{zichi2019}, optimized for saturation of internal efficiency and critical current which lead us to choose Nb$_{0.85}$Ti$_{0.15}$N and Nb$_{0.62}$Ti$_{0.38}$N as compositions for the VIS-NIR range and 1550 nm, respectively. The film thicknesses processed in this manuscript are 8 nm, 9 nm, 10.5 nm and 13 nm. The nominal thickness was estimated from the sputtering time and the deposition rate which has been verified by Atomic Force Microscopy. After film deposition, the wafers were cleaved in a small and large piece. The small pieces were used for determining the superconducting transition temperature in a four-point probe cryogenic setup and the large pieces were used to fabricate the detectors. \textbf{Figure \ref{fig:fig1}(a)} shows the normalized resistance versus base temperature for 8 nm and 13 nm thick films. In the inset, the table summarizes the critical temperature (Tc) for different film thicknesses studied in this work. We observe that the critical temperature increases with the film thickness. This behavior can be explained by the proximity effect \cite{Ilin_2008}, in addition to a decrease of the 3-dimensional disorder of the film \cite{PhysRevB.80.054510}. Subsequently, we used photolithography to deposit gold contact pads for the fiber-coupled devices. The meanders were patterned using the positive AR-P 6200 e-beam resist and a 100 keV electron beam lithography system, then the pattern was transferred to the NbTiN layer by dry etching using SF$_{6}$ and O$_{2}$ chemistry. Devices with different widths of nanowires were fabricated, from 50 nm to 100 nm, while keeping the filling factor (50 $\%$) and diameter (14 $\mu$m) constant. \textbf{Figure \ref{fig:fig1}(b)} presents a scanning electron microscope (SEM) image of a fabricated detector with the meander highlighted in red. The inset of \textbf{Fig \ref{fig:fig1}(b)} shows a magnified view of the detector with a nanowire width of 70 nm. To fiber couple the detectors, an additional etching step is needed to etch keyhole shaped chips\cite{Miller:11}. Next, the samples were mounted on two different printed circuit boards (PCB): \textbf{Fig \ref{fig:fig1}(c)} presents a picture of a flood illumination holder whereas \textbf{Fig \ref{fig:fig1}(d)} shows a fiber-coupling holder made from oxygen free copper. In both cases, a heater element which consists of a flat flexible resistor embedded in Kapton film (Iceoxford), is placed below the PCB to control the detector temperature. Moreover a cryogenic temperature sensor is glued with varnish in the proximity of the detector. Finally, the detectors were wire-bonded to the selected PCB for electrical bias and readout. We investigated the internal efficiency performance of devices made from different film thicknesses by using the flood illumination holder (\textbf{Fig \ref{fig:fig1}(c)}) as described in the first part of section 4. The measurements were performed with a home-made exchange gas dipstick immersed into a transportable liquid helium dewar.

\section{Measurements of SNSPDs critical current and dark count rate as a function of temperature}
\begin{figure}[hbtp!]
	  \centering
	  \includegraphics[width=13cm]{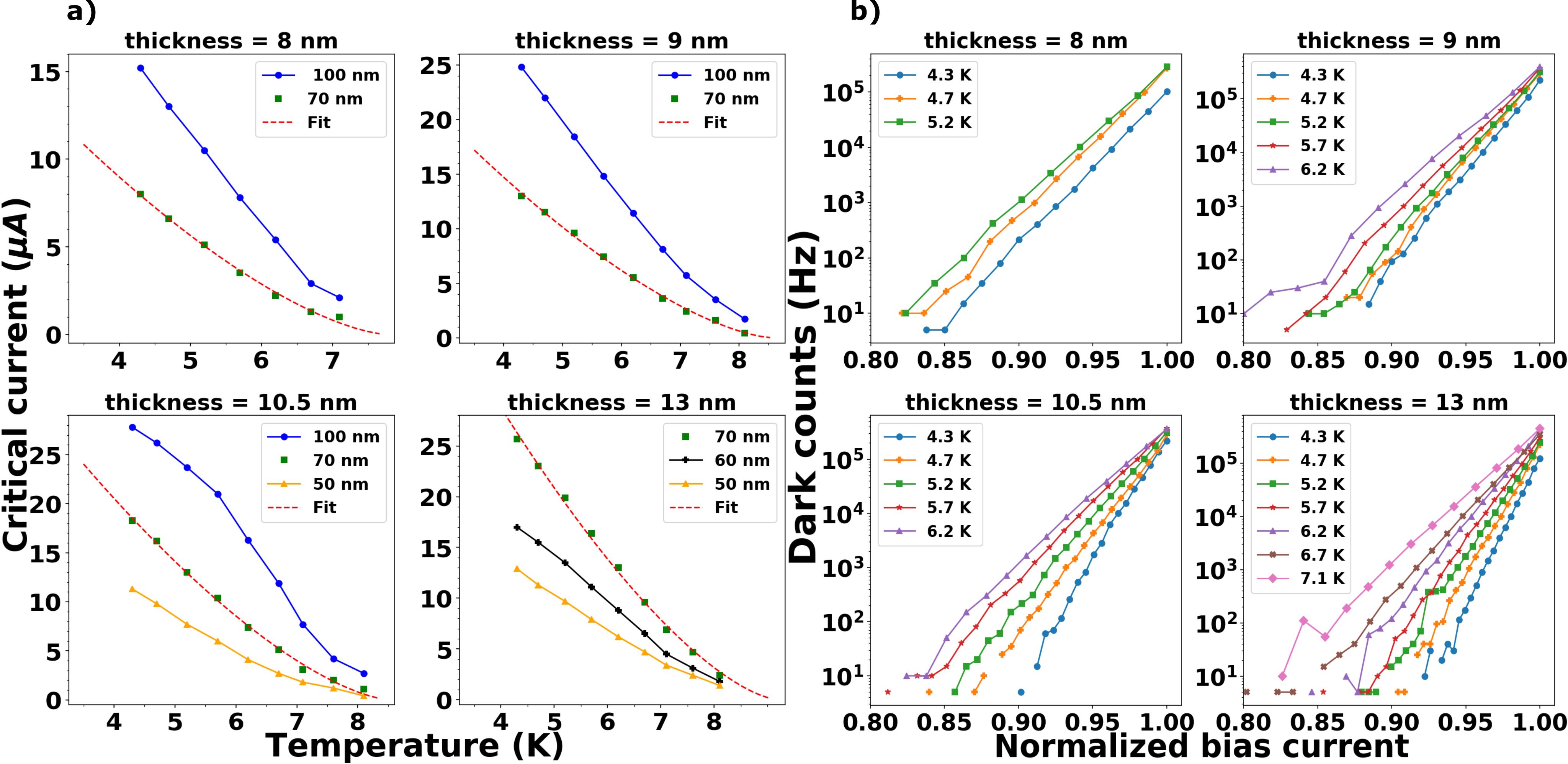}
	  \caption{(a) Measurements of the critical current vs. base temperature for different film thicknesses (8 nm, 9 nm, 10.5 nm and 13 nm) and nanowire width (50 nm, 60nm ,70 nm and 100 nm). The red dashed line is a fit from the Ginzburg-Landau theory.  
	           (b) Measurements of the dark count rates vs. bias current for different film thicknesses (8 nm, 9 nm, 10.5 nm and 13 nm) with 70 nm nanowire width.}
	\label{fig:fig2}
\end{figure}
                                                          
\textbf{Figure \ref{fig:fig2}(a)} shows the critical currents of SNSPDs made from films with different thicknesses and nanowire widths, operated at temperatures ranging from \SI{4.3}{K} to \SI{8.1}{K}. Above $\sim$ \SI{8}{K}, the critical current is too low to resolve a detection pulse from the noise (root mean square noise $\sim$ 30 mV) of our setup. We observe that the critical current decreases when the temperature increases, as expected. Additionally, the use of thicker films and wider nanowires increase the critical current of the superconducting devices \cite{doi:10.1063/1.3437043}. The red dashed curve in \textbf{Fig \ref{fig:fig2}(a)} is a fit from the Ginzburg-Landau theory (Ic $\propto$  \((1-\frac{T}{Tc})^\frac{3}{2}\)), which defines the behavior of the critical current as function of the temperature. For 70 nm wide nanowires, the fit leads to Tc= \SI{8.55}{K} and Tc= \SI{9.18}{K} instead of Tc=\SI{10.12}{K} and Tc= \SI{10.85}{K} that was measured before etching the superconducting film into a meander shape, for 9 nm and 12 nm film thickness, respectively. \textbf{Figure \ref{fig:fig2}(b)} presents the Dark Count Rate (DCR) of 70 nm wide nanowire SNSPDs as function of the bias current for different film thickness, in a temperature range from \SI{4.3}{K} to \SI{7.1}{K}. For these measurements, we used the flood illumination holder (\textbf{Fig \ref{fig:fig1}(c)}) in order to minimize the influence of black-body radiation that could be transmitted by the fiber and we assumed that only the intrinsic dark counts were recorded. We notice that the DCR raises exponentially with the bias current and the absolute value increases with the temperature. This behavior suggests that the dominant mechanism responsible for dark count is the current-assisted unbinding of vortex-antivortex pairs caused by thermal fluctuations \cite{doi:10.1063/1.3652908}. Moreover, for thicker films the exponential increases of DCR starts at higher bias current compared to the critical current. For instance, at \SI{4.3}{K} (blue curves in \textbf{Fig \ref{fig:fig2}(b)}) we can compare the starting bias current values for which the DRC curve starts to increase exponentially. It corresponds to a bias current of 0.85.Ic and 0.93.Ic for 8 nm and 13 nm film thickness, respectively. A similar trend is observed for all the investigated temperatures. To conclude this section, we have shown that it is possible to reduce the intrinsic DCR and increase the switching current (by up to a factor of 3 for 70 nm nanowire width) by using thick film for SNSPDs.

\section{SNSPDs efficiencies and timing jitter characterization as a function of temperature}

\begin{figure}[hbt!]
	  \centering
	  \includegraphics[width=13cm]{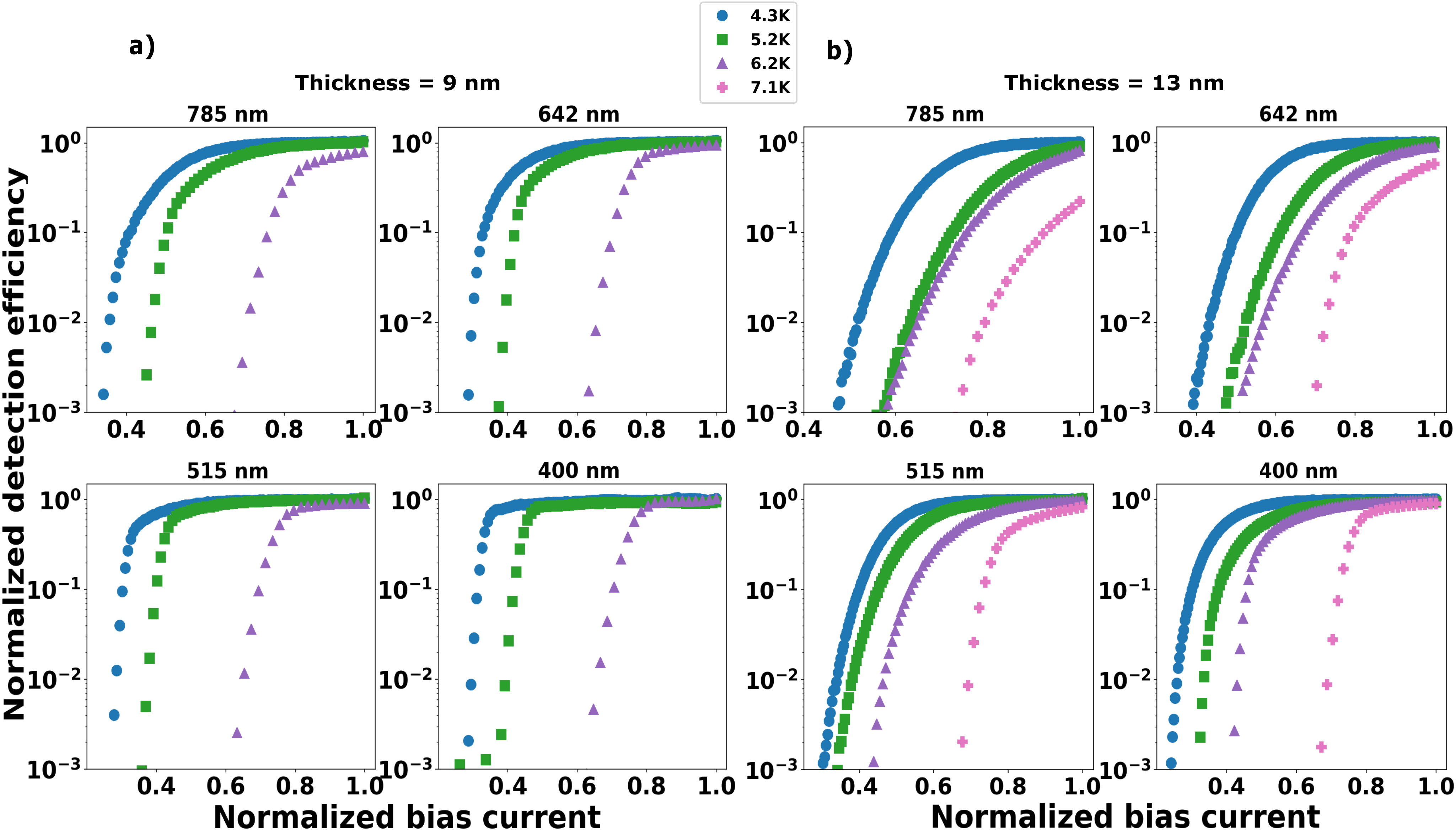}
      \caption{(a) Normalized detection efficiency vs. bias current at \SI{4.3}{K}, \SI{5.2}{K} and \SI{6.2}{K} for 9 nm thick film.
      (b) Normalized detection efficiency vs. bias current at \SI{4.3}{K}, \SI{5.2}{K}, \SI{6.2}{K} and  \SI{7.1}{K} for 13 nm thick film. The measurements are performed with CW laser diodes at 785 nm, 642 nm, 515 nm and 400 nm. 
		        }
	\label{fig:fig3}
\end{figure}

To characterize the single photon detectors, we used different laser diodes emitting at 400 nm, 515 nm, 642 nm, 670 nm, 785 nm and 1550 nm. The wavelength range of 400 - 700 nm is significant for fluorescence microscopy \cite{doi:10.1002/open.201700177,Rockland} and for studying celestial objects from space and ground-based observatories. Many high purity III-V QDs emit at $\sim$ 800 nm \cite{doi:10.1063/1.5020038,SchAll2019} likewise quantum memories based on electromagnetically induced transparency in Rubidium cells \cite{PhysRevLett.100.093602}. 1550 nm wavelength is the widely used wavelength for quantum communication technology.  
\\
\textbf{Figure \ref{fig:fig3}(a)} presents the normalized detection efficiency of the device as a function of the bias current for 70 nm wide nanowire and 9 nm film thickness, from \SI{4.3}{K} to \SI{6.2}{K}. Each data point is an average of three measurements with an integration time of 200 ms, and the DCR is subtracted. From \SI{4.3}{K} to \SI{6.2}{K}, we observe a strong saturation of the internal efficiency for all wavelengths, except for 785 nm at \SI{6.2}{K} where the saturation is weak. By increasing the film thickness, we enhance the optical absorption for longer wavelengths but the internal efficiency gets weaker \cite{doi:10.1063/1.5000001}. \textbf{Figure \ref{fig:fig3}(b)} presents the normalized detection efficiency from \SI{4.3}{K} to \SI{7.1}{K} as function of the bias current, for 70 nm wide nanowire and 13 nm film thickness. Similar to 9 nm thickness, the saturation of the internal efficiency is reached from \SI{4.3}{K} to \SI{6.2}{K}, but decreases with higher photon wavelength. At \SI{7.1}{K}, the saturation regime is attained for 400 nm and 515 nm. As expected, at low temperatures (\SI{4.3}{K}) devices made from the thinner 9 nm films exhibits stronger saturation of internal efficiency than devices from the thicker 13 nm films. However, at \SI{7.1}{K}, 13 nm film thickness still demonstrate saturation for 515 nm and 400 nm.  The electrical output signal from 9 nm film is to weak to expose any saturation behavior. Therefore, at higher temperatures 13 nm film outperforms 9 nm film.

\begin{figure}[hbtp!]
	  \centering
	  \includegraphics[width=13cm]{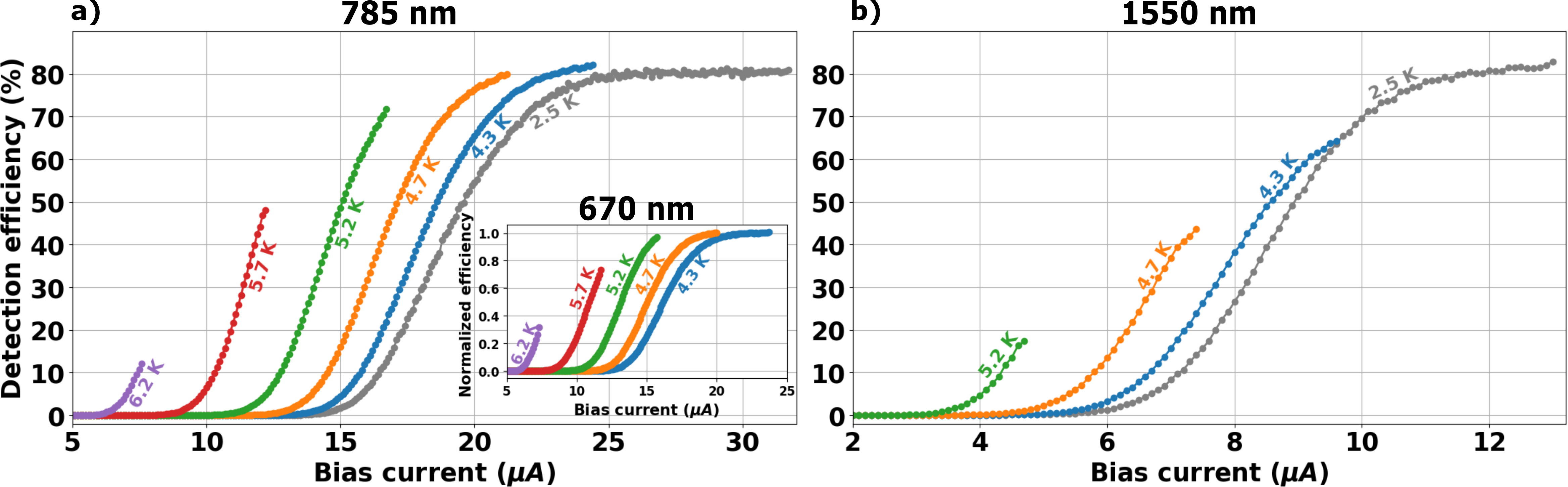}
      \caption{(a) Detection efficiency measurements at 785 nm for T= 2.5 - \SI{6.2}{K} vs. bias current. The inset presents the normalized detection efficiency at 670 nm vs. bias current in the same temperature range.
       (b) Detection efficiency measurements at 1550 nm for T= 2.5 - \SI{5.2}{K} vs. bias current.}
	\label{fig:fig4}
\end{figure}
Finally, we fabricated fully packaged devices in order to measure the efficiencies at $\sim$ 800 nm and 1550 nm as well as the temporal resolution at higher base temperatures. The fiber-coupled lasers were attenuated with a NIST-traceable attenuator to a level corresponding to a photon flux of $\sim$ 80 kphotons/s. The light polarization was set by using a fiber paddle polarization controller to the transverse electric mode such that the SNSPD has maximum optical absorption. The integration time for measuring the efficiency is 1 s and the DCR is subtracted. The main contribution of error in our measurements originated from the power meter, 3 $\%$ for 785 nm and 5 $\%$ for 1550 nm, while the laser power is stable within 1 $\%$ for both wavelength. Thus, we estimated the measurement error to be 4 $\%$ for 785 nm and less than 6 $\%$ for 1550 nm. \textbf{Figure \ref{fig:fig4} (a)} shows the detection efficiency at 785 nm in the temperature range of \SI{2.5}{K} to \SI{6.2}{K}. The inset is the normalized efficiency at 670 nm for the same device, in the temperature range of \SI{4.3}{K} to \SI{6.2}{K}. The measurements at \SI{2.5}{K} display in grey in \textbf{Fig\ref{fig:fig4} (a)} and \textbf{Fig\ref{fig:fig4} (b)} were carried out in a Gifford-McMahon closed cycle system. At \SI{2.5}{K} in \textbf{Fig\ref{fig:fig4} (a)}, the detection efficiency starts to saturate at 25.5 $\mu$A and reaches 80 $\%$. By considering the fiber-air interface reflection during measuring the input power, the system detection efficiency is 77 $\%$. At liquid helium temperature, the saturation plateau gets weaker but the detection efficiency is maintained and yield to 82 $\%$. The difference in detection efficiency between \SI{2.5}{K} and \SI{4.3}{K}, which should be identical, is attribute to the use of different sets up for the measurements. Until \SI{5.2}{K}, the detection efficiency remains above 70 $\%$  and reaches 15 $\%$ at \SI{6.2}{K}. At 670 nm the saturation behavior is maintained until \SI{5}{K}.
\textbf{Figure \ref{fig:fig4} (b)} shows the detection efficiency at 1550 nm in the temperature  of \SI{2.5}{K} to \SI{5.2}{K}. At \SI{2.5}{K}, the detection efficiency plateaus to 81 $\%$. Similar to 785 nm, at \SI{4.3}{K} the saturation gets weaker and reaches 64 $\%$ of detection efficiency. For \SI{4.7}{K} and \SI{5.2}{K} the detection efficiency is above 40 $\%$ and 15 $\%$, respectively.

\begin{figure}[hbtp!]
	  \centering\includegraphics[scale=0.6]{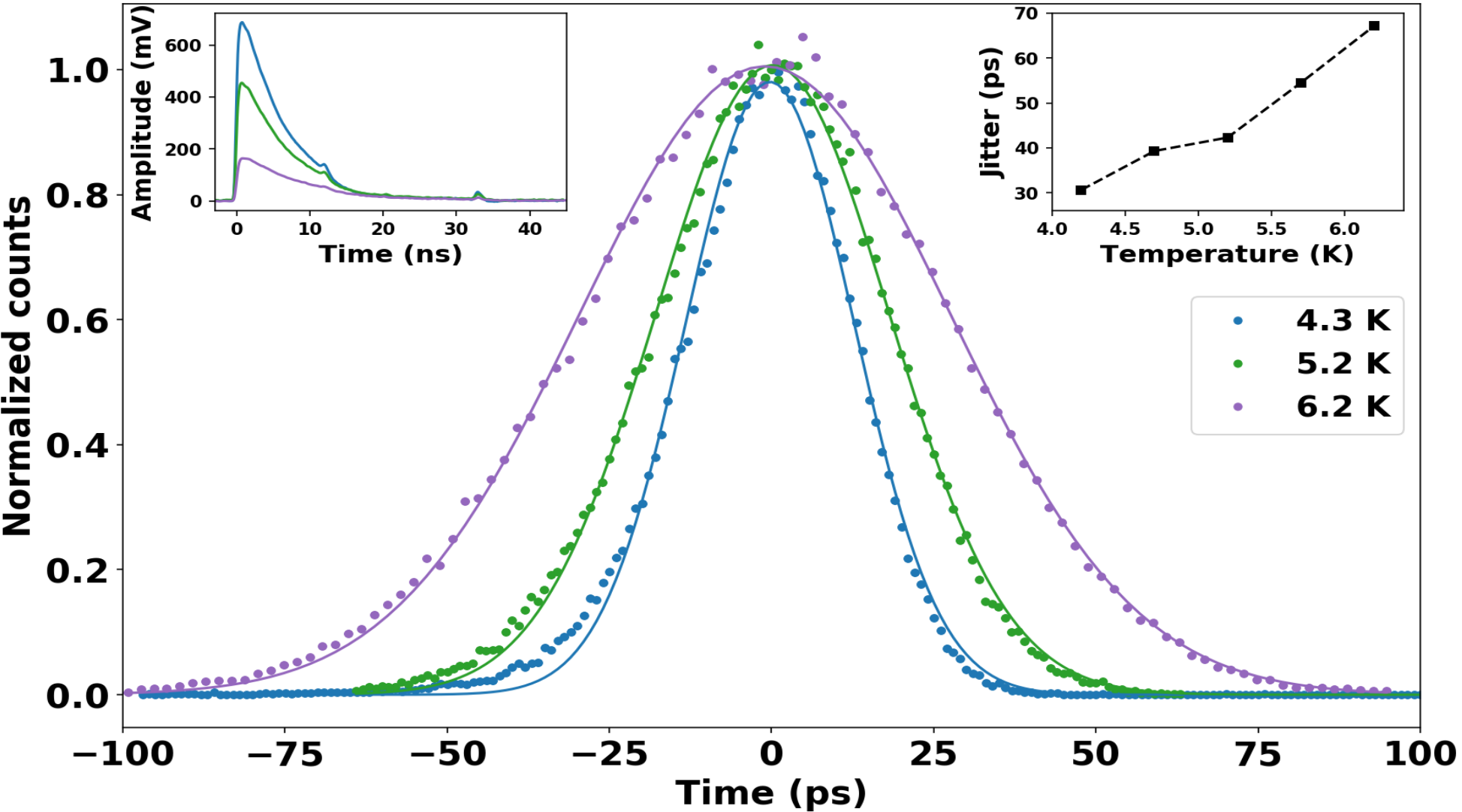}
      \caption{Timing jitter measurements (dots) and their corresponding fits (lines) at base temperature of \SI{4.3}{K}, \SI{5.2}{K} and \SI{6.2}{K}. The left inset is the detection pulse of the SNSPD at base temperature of \SI{4.3}{K}, \SI{5.2}{K} and \SI{6.2}{K}. The right inset is the timing jitter as a function of the temperature.}
	\label{fig:fig5}
\end{figure}

The timing jitter was measured as function of the base temperature and to get an optimal result, we replaced the room-temperature amplification with a cryogenic amplifier cooled down to 77 K. We used a picosecond pulsed laser (center wavelength of 1064 nm, 50 MHz repetition rate) and an fast digital oscilloscope (Lecroy Waverunner 640Zi 4 GHz,40 GS/s) as correlator. \textbf{Figure \ref{fig:fig5}} presents the timing jitter measurement results of the detector biased at 90 $\%$ of its critical current at the temperatures of \SI{4.3}{K}, \SI{5.2}{K} and \SI{6.2}{K}. The signal to noise ratio decreases with temperature which results in a deterioration of the temporal resolution of the detector \cite{doi:10.1063/1.5000001}. 
From \SI{4.3}{K} to \SI{6.2}{K}, the amplitude of the electrical output pulse of the superconducting device is lowered by a factor $\sim$ 3, as shown in the left inset of \textbf{Fig \ref{fig:fig5}}. The right inset of \textbf{Fig \ref{fig:fig5}} presents the timing jitter of the detector depending on the base temperature. The fitted data gives a full width half maximum (FWHM) timing-jitter of 30.6 $\pm$ 0.3 ps and 67.2 $\pm$ 0.3  ps at \SI{4.3}{K} and \SI{6.2}{K}, respectively.

\section{Conclusion}
In summary, we have demonstrated that NbTiN superconducting material is suitable to operate in cryostats with base temperatures as high as \SI{7}{K}. By optimizing the film thickness and nanowire width of the superconducting single photon detector, we showed a detection efficiency of 82 $\%$ at 785 nm combined with a temporal resolution as high as 30.6 $\pm$ 0.3 ps at \SI{4.3}{K}. For a similar operating temperature, we measured a detection efficiency of 64 $\%$ at 1550 nm. Moreover, we reached saturation of the internal efficiency for the whole visible spectrum, up to \SI{7}{K} while preserving a high critical and by increasing the film thickness of the superconducting device, we decrease the intrinsic dark count rate. The devices reported in this work paves the way for integration of large arrays of high performance SNSPDs in emerging or well establish applications such as astronomy, quantum applications and bio-imaging.

\section*{Funding}
This work was supported by the European Commission via the Marie-Sklodowska Curie action Phonsi (H2020-MSCA-ITN-642656). I.E.Z. acknowledges the support
of NWO LIFT-HTSM for Physics 2016-2017, project no. 680-91-202.

\section*{Acknowledgments}
We would like to thank Dr. M.J.A. de Dood for valuable discussions and A. W. Elshaari for his help with the publication.

\bibliography{Reference_HTc}

\end{document}